\documentclass{JHEP3}
\usepackage{epsfig,amsmath,amsthm,latexsym}
\newcommand{\be}{\begin{equation}}
\newcommand{\ee}{\end{equation}}
\newcommand{\bea}{\begin{eqnarray}}
\newcommand{\eea}{\end{eqnarray}}

\def\IZ{\relax\ifmmode\hbox{Z\kern-.4em Z}\else{Z\kern-.4em Z}\fi}
\newcommand{\IS}{{\bf S}}

\newcommand{\non}{\nonumber \\}

\def\half{{\frac12}} 
\def\tr{{\rm tr}}
\def\del{{\partial}}

\def\tr{\tilde{r}}

  \def\eps{\epsilon}

\def\presub{\vspace{.5cm} \noindent}

\def\bi{\begin{itemize}} \def\ei{\end{itemize}}

\def\Schw{Schwarzschild }
\def\({\left(} \def\){\right)}
\def\[{\left[} \def\]{\right]}

\newcommand{\lam}{\lambda} \newcommand{\Lam}{\Lambda}

\def\dh{\hat{d}}
\def\rh{\hat{r}} \def\xh{\hat{X}}
\def\rt{\tilde{r}}
 \def\lamt{\tilde{\lambda}} \def\lamb{\bar{\lambda}}
\preprint{WU-AP/267/07}

\title{ \center{High and Low Dimensions in The Black Hole Negative Mode}}

\author{Vadim Asnin\footnotemark[1], Dan Gorbonos\footnotemark[1],
 Shahar Hadar\footnotemark[1], Barak Kol\footnotemark[1], Michele
 Levi\footnotemark[1] \hspace{3cm}
 and Umpei Miyamoto\footnotemark[2] \\

\footnotemark[1]
 Racah Institute of Physics \\
 The Hebrew University \\
 Jerusalem 91904, Israel \\
{\tt \href{mailto:vadima@pob.huji.ac.il}{vadima@pob.huji.ac.il}, \\
  \href{mailto:gdan@phys.huji.ac.il}{gdan},
 \href{mailto:shaharhadar@phys.huji.ac.il}{shaharhadar},
   \href{mailto:barak_kol@phys.huji.ac.il}{barak\_kol},
   \href{mailto:michele@phys.huji.ac.il}{michele} @phys.huji.ac.il}
\\

\footnotemark[2] Department of Physics\\
 Waseda University, Okubo 3-4-1\\
  Tokyo 169-8555, Japan\footnotemark \\
 \email{umpei@gravity.phys.waseda.ac.jp}}

\abstract{The negative mode of the Schwarzschild black hole is
central to Euclidean quantum gravity around hot flat space and for
the Gregory-Laflamme black string instability. We analyze the
eigenvalue as a function of space-time dimension $\lam=\lam(d)$ by
constructing two perturbative expansions: one for large $d$ and the
other for small $d-3$, and determining as many coefficients as we
are able to compute analytically. Joining the two expansions we
obtain an interpolating rational function accurate to better than
2\% through the whole range of dimensions including $d=4$.}

\begin{document}

\section{Introduction and Summary}

The \Schw black hole is known to possess a single off-shell negative
mode\footnote{This mode does not represent a physical instability of
the black hole but rather of the black string constructed from it.}
discovered by Gross, Perry and Yaffe (GPY) \cite{GPY}. It plays a
central role in Euclidean Quantum Gravity around hot flat space
\cite{GPY} and in the Gregory-Laflamme (GL) black string instability
\cite{GL1,rev,HOrev2}.

In the context of hot flat space the Euclidean black hole appears as
a saddle point of the action with the same asymptotics, namely a
periodic Euclidean time direction, and is interpreted to represent a
non-perturbative (tunneling) decay mode through the nucleation of a
real black hole. The resulting leading correction to the spacetime
energy density comes from the exponential of minus the action. If
this non-perturbative effect is to describe an instability this
correction should be imaginary and whether this is the case is
determined by the subleading correction given by the square-root of
the perturbations determinant. Therefore Gross-Perry-Yaffe argue
that the consistency of this picture requires an odd number of
negative modes, and in particular at least one mode should exist.

The Gregory-Laflamme string instability occurs only for
wavenumbers smaller than a certain critical wavenumber $k_{GL}$,
which translates into a critical mass in the presence of a fixed
compactification length. $k_{GL}$ has a clear physical meaning and
it is directly determined by the negative eigenvalue, as the
latter is nothing but $(-k_{GL}^2)$.

The negative eigenvalue was computed numerically in 4d to be
\cite{GPY} $\lambda(4d) \simeq 0.76$
and extended to arbitrary dimensions in
\cite{SorkinD*,KolSorkin}\footnote{The data was obtained in
\cite{SorkinD*} and published in \cite{KolSorkin}.} where it was
shown that the behavior of $k_{GL}(d)$ brings insight into the
discovery of a critical dimension where the GL transition turns
second order \cite{SorkinD*}.\footnote{More insight into this
critical dimension and other aspects of the problem can be gleaned
from a hydrodynamical analogue \cite{CardosoDias} of the
Rayleigh-Plateau instability which is responsible for the creation
of drops in a faucet.} The negative eigenvalue was also computed
in \cite{GL1} for $4 \le d \le 9$, in \cite{Wiseman2002} for $5d$,
and in \cite{KudohMiyamotoSmeared} Harmark-Obers coordinates were
used to obtain a master equation. The computation was generalized
in \cite{HovdeboMyers} to boosted strings, in
\cite{KudohMiyamotoCharged} to charged strings and in \cite{KKR}
to rotating strings. Finally related non-uniform strings in
various $d$ appeared in \cite{SorkinVariousD}.

While an analytic determination of $\lambda(d)$ is not known
 it is usually useful to have some analytic
insight. In this paper we attempt for analytic control by a
standard strategy in physics -- looking at extreme limits and
perturbing around them with a small parameter. Here, taking $d$ as
a variable allows for two limits: high $d$ and low $d$ where for
reasons to be described later at low $d$ we are interested in $d
\to 3^+$ (approaching 3 from above). Actually the large $d$ limit
of this problem was already considered in \cite{KolSorkin} but
here we are able to obtain more results as discussed towards the
end of this introduction.

In high dimensions gravity becomes increasingly short-ranged, and
this limit was already considered as a simplified limit for the
Feynman diagrams of quantum gravity
\cite{Strominger-HighD,Bohr-HighD,LargeD-QG}. The idea of small,
continuously varying dimensions is also familiar from the technique
of Dimensional Regularization in field theory \cite{DimReg}.
Moreover, 3d gravity is known to have an interesting and somewhat
degenerate limit (see \cite{3dGrav},\cite{Deser} and references
therein) where gravitational waves do not exist and space-time is
flat apart from deficit-angle point singularities at the location of
each particle.

Recently ``the optimal gauge'' master eigenvalue equation for the
negative mode was obtained in \cite{nGPY,LG-GL}. This ``optimal
gauge'' was generalized by proving that for perturbations of any
spacetime with at most one non-homogeneous coordinate the gauge
freedom can always be eliminated \cite{1dPert} thus avoiding the
need to choose a gauge fix. The latter general theorem was already
applied to a re-derivation of gravitational waves in the \Schw
background \cite{nRW}.

In this paper we utilize the new master equation to analytically
study the function $\lam(d)$ for both large and small dimensions.
In section \ref{setup-section} we set the stage by introducing the
master equation and showing that in both the high and low $d$
limits it can be recast in terms of two different variables and
analytic control of $\lam$ can be achieved through the technique
of Matched Asymptotic Expansion (MAE) between the near horizon and
asymptotic regions.
In sections \ref{highDsection} and \ref{lowDsection} we analyze
the equations for high and low dimensions respectively. In section
\ref{InterSection} we compare the analytic results of the previous
two sections to new high-precision numerical data (summarized in
appendix \ref{numeric}) and propose an interpolating rational
function valid over the whole range of $d$. The interpolation
resembles the Pad\'{e} approximation only here it is based on two
Taylor expansions (rather than one) from both ends of the range $3
\le d \le \infty$.

\presub {\bf Summary of results.} The high $d$ expansion is found to
be \be
 \lam = \dh - 1 + \frac{2}{\dh} + \dots \equiv d - 4 + \frac{2}{d} + \dots ~, \label{highd-sum} \ee
 where $\dh \equiv d-3$. 
 See a discussion of the reliability of the last term  of (\ref{highd-sum}) in the second paragraph of section
 \ref{InterSection}.
 The low $d$ expansion is \bea
 \lam &=& c_1\, \eps + c_2\, \eps^2 \dots \equiv c_1\, (d-3) + c_2\, (d-3)^2 \dots \non
 c_1 &\simeq& 0.71515 \non
 c_2 &\simeq& 0.0627  ~, \label{lowd-sum} \eea
 where $\eps \equiv \dh \equiv d-3$ is small, and while the leading ${\cal O}\(\eps\)$ behavior
was inferred analytically the constants $c_1,\, c_2$ were
determined numerically.

The two expansions can be combined into the following rational
approximation valid throughout the range $0 \le \dh \le \infty$ \be
 \lam(d) \simeq \dh ~ \frac{(1-c_1)\, \dh + c_1}{(1-c_1)\, \dh +
 1} ~. \label{interpol1}
 \ee
The approximation was built to match both series (apart for the
last term of each one), and it is found to uniformly agree with
the numerical data to better than 2\%. In particular by
approximating $c_1 \approx 0.7$ we find the following handy
formula \be
 \lam \simeq \dh\, \frac{3\, \dh + 7}{3\, \dh + 10} ~, \label{interpol2} \ee
  and in 4d \be
 \lam(4d) \approx \frac{10}{13} = 0.769 ~,\ee
 which is in good agreement with our available numerical value $\lam(4d)=0.76766$.

\presub {\bf Discussion}. In general the idea to study GR in various
dimensions, high and low, is received with considerable resistance
from both laymen and professionals. In our opinion, one of the
important motivations to consider various $d$'s is the view that
General Relativity is defined for all $d$, and hence $d$ should be
considered as its parameter. From this point of view it is as
natural to study GR is various $d$'s as it is to study field
theories with matter content that differs from the standard model.
We view the success of the analysis presented here, including the
insight into 4d results as a vindication of this ``dimension as a
parameter'' approach. For completeness it should be mentioned that
there are at least two other important reasons to study GR in
various $d$'s: theoretical reasons from String Theory, and
``phenomenological'' reasons in the context of the Large Extra
Dimensions and braneworld scenarios.

{\it Comparison with \cite{KolSorkin}}. In \cite{KolSorkin} the
leading order high $d$ behavior of $\lambda(d)$ was sought
analytically. Here we are able to go further, getting two
subleading orders as well as finding the low $d$ behavior. Another
difference is that the ``optimal gauge'' master equation was not
available at the time of \cite{KolSorkin}, but rather the
transverse-traceless gauge was used, adding a spurious singularity
to the equation and presumably obstructing the analysis in the
near zone.

{\it Open questions}. It is plausible that our interpolating
formulae (\ref{interpol1},\ref{interpol2}) can be further refined.
For instance $\lam \simeq (d-1)(d-3)/d$ is interesting.
Generalization to other perturbation problems is to be
expected.\footnote{For quasi normal modes some high $d$ behavior is
already known \cite{Konoplya}.} Another point is that since the
Matched Asymptotic Expansion analysis amounts to forgetting the
asymptotic region altogether in the actual calculation, it is
possible that there is a more elegant argument to demonstrate this
fact.

{\bf Note added} (v3). While our results are unchanged we added
 qualifying remarks on the assumed decoupling of the asymptotic
 region from the near horizon region (for both high and low $d$).

\section{Set-up}
\label{setup-section}

The metric of the $d$-dimensional \Schw black hole in standard
Schwarzschild coordinates is given by \bea
 ds^2 &=& -f\, dt^2 + f^{-1}\, dr^2 + r^2\, d\Omega^2_{d-2} \non
  f(r) &=& 1- \( \frac{r_0}{r} \)^{d-3} ~, \label{schw-coord} \eea
 where $r_0$, the \Schw radius, is the location of the horizon and
$d\Omega^2_{d-2}$ denotes the metric of the round $d-2$ sphere
$\IS^{d-2}$. Hereafter we shall use units such that $r_0=1$.

The ``optimal gauge'' master equation for the negative mode is
\cite{nGPY,LG-GL} \bea
 & & \[ -\frac{1}{r^{d-2}}\, \del_r\, f\, r^{d-2}\, \del_r + V(r) \]
 \psi = -\lam\, \psi \non
 & & V(r) := - \frac{2(d-1)(d-3)^3}{r^2\(2(d-2)r^{d-3}-(d-1)\)^2} ~,
 \label{master}
 \eea
where $\psi$ is a master field which is a gauge invariant
combination of the metric perturbation components
$h_{tt},h_{\Omega\Omega}$ (see \cite{nGPY} for the full definition).
Our sign conventions were chosen such that a negative eigenvalue is
represented by a positive $\lam$.

Note that the eigenvalue $\lam=\lam(d)$ is related to the critical
Gregory-Laflamme wavenumber for the $\mathit{d+1}$ dimensional black
string through \cite{GL1} \be
 \lam = k_{GL}^2 ~.\ee

Large $d$ is clearly an interesting limit. Numerical data suggests
that in 3d $\lam$ vanishes. $d=3$ is also seen to be special from
the definition of $f(r)$ (\ref{schw-coord}). Therefore we shall
attempt to analyze also $d \to 3^+$. Altogether our two limits are
\bi
 \item High $d$, namely $d \to \infty$ where the small
parameter is $1/d$
 \item Small $d$, namely $d \to 3^+$ where the small parameter is
 $d-3$.
 \ei
Accordingly we find it useful to denote \be
 \dh := d-3 \label{def-dh} ~. \ee

The appearance of the term $r^{d-3}$ in the definition of $f$
(\ref{schw-coord}) and in several other places in the master
equation (\ref{master}) suggests to define the variable \be
 X := r^{\dh} \equiv r^{d-3} \label{def-X} ~.\ee
 The master equation turns out to have a nice representation in
the $X$ variable \bea
 & & \[ - \del_X X(X-1)\, \del_X + V_X(X) \]
 \psi = -\frac{r^2 \lam}{\dh^2}\, \psi \non
 & & V(X) := - \frac{2(\dh+2)\dh}{\(2(\dh+1)\,X-(\dh+2)\)^2} ~,
 \label{Xmaster}
 \eea
gotten by multiplying (\ref{master}) by $r^2/\dh^2$, and here $r$
should be considered to be a function of $X$.

It will also be convenient to define \be
 \rh:=r-1 \qquad \xh := X-1 ~, \label{def-rxh} \ee
 such that the horizon lies at $0=\rh=\xh$.

In both high and low $d$ limits we can define two regions (or
zones) on the $r,X$ axis namely, the near horizon region and the
asymptotic region, and moreover their overlap will increase in
size as the limit is taken. According to the method of matched
asymptotic expansion, one solves for $\lam=\lam(d)$ as follows.
One solves for $\psi=\psi(r;\lam)$ separately in each region each
time with only a single boundary condition - regularity at the
boundary. The second boundary condition comes from matching in the
overlap zone. Normally when propagating a wavefunction from two
boundaries one requires $\psi'/\psi$ to be continuous at the
matching point. When we have a matching region, rather than a
point, $\psi$ can be characterized as a linear combination of two
basic solutions, namely we denote the general solution inside the
overlap region (which is usually simpler) $\psi= A(\lam)\,
\psi_1(r;\lam) + B(\lam)\, \psi_2(r;\lam)$ where $A,\, B$ are two
arbitrary coefficients. The matching condition becomes \be
 \left. \frac{A}{B}\right|_{\rm near}(\lam) = \left. \frac{A}{B}\right|_{\rm
 asymp}(\lam) ~, \ee
 where $A,B$ from each region need to be read from the asymptotics
of $\psi_{\rm region}(r;\lam)$ \emph{away} from the boundary. This
is the condition that fixes the eigenvalue $\lam$.

Actually in the two limits under consideration, matching will not
be necessary: we shall start the next two sections by
demonstrating that in both cases it is enough to analyze the near
horizon region and to replace the matching boundary conditions by
a second regularity condition in the far side of the near horizon
region zone. This means that both wave functions are localized in
the near horizon region. \footnote{Note that there are examples
where this simplification does not occur. For example, in the 1d
Quantum Mechanics problem of finding a bound state for a shallow
potential, $[-\del_x^2 + \eps\, V(x)] \psi= E(\eps)\, \psi$ the
wave function of the bound state extends much beyond the size of
the potential.}

\newpage
\section{High $d$}
\label{highDsection}

\begin{figure}[t!] \centering \noindent
\includegraphics[width=5cm,angle=-90]{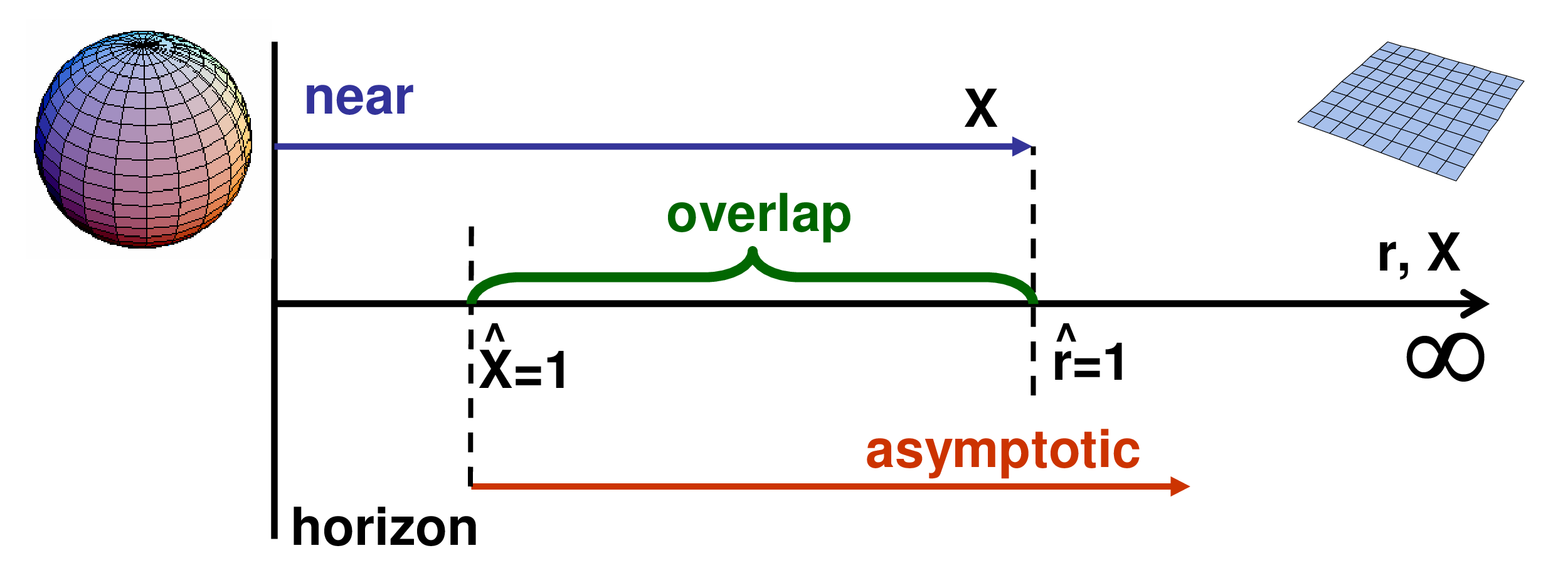}
\caption[]{The regions along the radial direction at high
dimension.}\label{high-figure}
\end{figure}

For high $d$ we have $\rh \ll \xh$  and hence $\xh=1$ occurs much
before $\rh=1$ on the $r,X$ axis (see figure \ref{high-figure}).
Accordingly we can define two regions \bi
 \item Near horizon $\rh \ll 1$
 \item Asymptotic $\xh \gg 1$
 \ei
and the two regions have an overlap which increases in size as the
large $d$ limit is taken, in agreement with the general
requirements of the method of matched asymptotic expansion.

\presub {\bf Asymptotic region and boundary conditions for  the
near horizon region}.
%
In the asymptotic region $f \simeq 1$ and $V \simeq 0$ and hence
the background is simply flat $d$-dimensional space-time. In this
region the master equation becomes \be
 -\frac{1}{r^{d-2}}\, \del_r\,  r^{d-2}\, \del_r\,
 \psi = -\lam\, \psi ~.
\ee $\lam$ can be rescaled out of this equation after defining the
dimensionless coordinate $\rt:=\sqrt{\lam}\, r$, and the solution
satisfying regularity at infinity is \be
 \psi = \frac{K_{\dh/2}(\tr) }{\tr^{\dh/2}} ~,\label{Bessel} \ee
 where $K_{\dh/2}$ is the modified Bessel function of the second kind of order $\dh/2$.

\noindent {\it Overlap region}. In the overlap region both $\rh \ll
1$ and $\xh \gg 1$, the $\lam$ term in the equation becomes
negligible and hence both $r$ and $X$ equations
(\ref{master},\ref{Xmaster}) become \be
 -\frac{1}{\rt^{d-2} }\, \del_{\rt} \,  \rt^{d-2}\, \del_{\rt} \,
 \psi = 0 = - \del_X\, X^2\, \del_X\, \psi  ~, \non
 \ee
 whose general solution is  \bea
 \psi &=& A\, + B\, \frac{1}{X} \non
      &=& A + B \frac{1}{r^{\dh}} ~. \eea


By examining the near horizon behavior of the Bessel function
(\ref{Bessel}) we convinced ourselves that to leading order
 \be
 A=0 ~.\label{highd-bc} \ee
Below we shall adopt a working assumption that this boundary
condition (which implies that the near horizon region does not
receive ``communications'' from the asymptotic region) is correct
to all order which we compute. We view the good agreement of the
analytic and numerical results as a confirmation of this
assumption (see the second paragraph of section
\ref{InterSection}).

\presub {\bf The eigenvalue problem}. In the near horizon region
$X$ is a good variable
 \footnote{An alternative change of variables which is of certain convenience is
$\xi:=-\tanh(\log(f)/2)=1/(2 X-1)$.}
 and we rewrite the $X$ master
equation (\ref{Xmaster}) in a form convenient for the high $d$
limit \be
 \[ - \del_X X(X-1)\, \del_X + V(X) \]
 \psi = -r^2(X)\, \lamt \, \psi ~,\ee
where the rescaled eigenvalue is \be
 \lamt:= \frac{\lam}{\dh^2} ~, \ee
while the potential and $r^2(X)$ are given by \bea
 V(X) &=& -\frac{2 \( 1+2/\dh \)}{\(2X-1+ 2(X-1)/\dh\)^2}\\
 r^2 &=& \exp\(\frac{2}{\dh}\,\log X \) ~.\eea

 \noindent {\it Zeroth order}. The zeroth
order equation in the near horizon region as $\dh \to \infty$ is
\be
 L_h\, \psi = -\lamt\, \psi \label{0th} ~, \ee
 where the horizon zeroth order differential operator is given by
 \be
 L_h := -\del_X\, X (X-1)\, \del_X - \frac{2}{(2 X-1)^2} \label{def-Lh} ~.\ee

The boundary conditions for this equation are regularity both at
the horizon $\xh=0$ and asymptotically at $\xh \to \infty$. More
specifically, near the horizon the two independent solutions
behave as $const, \log(\xh)$ and we require that the coefficient
of the $\log(\xh)$ piece vanishes. Asymptotically the two
solutions behave as $const,\, 1/\xh$ and we require that the
constant vanishes.

The zeroth order equation (\ref{0th}) has a \emph{marginally bound
state}, namely a normalizable solution with $\lamt=0$ given simply
by \be
 \psi_0=\frac{1}{2 X-1} \label{psi0} ~.\ee
 Indeed this solution satisfies the boundary conditions both at the horizon and at
 infinity and it has finite norm \be
 \left< \psi_0|\psi_0 \right> = \half \ee
under the inner product which makes $L_h$ Hermitian. This inner
product is nothing but the standard $L^2$ norm \be
 \left< \phi|\psi \right> = \int_1^\infty \phi^*(X)\,
 \psi(X)\, dX ~.
\ee

Since $\psi_0$ has no nodes we also conclude that it is the ground
state, and hence $\lamt=0$ at ${\cal O}(\dh^0)$.

We proceed by the standard perturbation theory for an eigenvalue
problem. At each order $k$ we find an equation of the form \be
 L_h ~ \psi_k = - \lamt_k\, \psi_0 + Src_k \label{pert} ~,\ee
 where $\lamt_k,\, \psi_k$ are the coefficients in the $1/\dh$ expansion
  of the eigenvalue and eigenfunction, respectively \be
  \lamt = \sum_{k=0}^{\infty} \lamt_k/\dh^k \qquad \psi = \sum_{k=0}^{\infty}
  \psi_k/\dh^k  ~.\ee
$Src_k$ is the source term at the $k$'th order which comes from the
$1/\dh$ expansion of (\ref{Xmaster}) and involves $\lamt$ and $\psi$
from lower orders. More specifically one needs to expand in
(\ref{Xmaster}) the near horizon potential, $V(\xh)$, and
$r^2=X^{2/\dh}=\exp(2 \log(X)/\dh)$.

Equation (\ref{pert}) is solved first by taking the inner product
with $\psi_0$ to find
 \footnote{Note that the boundary condition
(\ref{highd-bc}) was used to derive this, and that a different one
could have required to add a boundary term.}
\be
 \lamt_k = \frac {\left< \psi_0\, | Src_k \right>}{\left< \psi_0\, | \psi_0
 \right>} \label{solve-lam} ~.\ee
 Next $\psi_k$ is solved from the inhomogeneous differential
equation (\ref{pert})  while imposing regularity at the boundaries.
This task is facilitated by the existence of a simple solution
(\ref{psi0}) to the homogenous equation and by the use of the
Wronskian. At each order there is a residual freedom to add to
$\psi_k$ any multiple of $\psi_0$ which amounts to a $\dh$-dependent
normalization factor for $\psi$. While it could be fixed by
requiring $\psi_k$ to be orthogonal to $\psi_0$, it is not necessary
to do so, and we use this freedom to simplify the expression for
$\psi_k$ when possible.

\presub {\it First order}. The first order correction to the
potential is \be
 V_1 = -\frac{4}{(2 X - 1)^3} \ee
 and the first source term is \be
 Src_1 = -V_1\, \psi_0  ~.
\ee

 Hence by (\ref{solve-lam}) the first correction to $\lamt$ is \be
 \lamt_1 = -\frac{\left< \psi_0|V_1|\psi_0 \right>}{\left< \psi_0|\psi_0
 \right>}=1 ~.\ee
 Solving (\ref{pert}) for $\psi_1$ we find \be
 \psi_1 = \frac{1}{(2X-1)^2}-\frac{\log X}{2X -1} ~.
 \ee

\presub {\it Second order}. The second order correction to the
potential is \be
 V_2 = +\frac{8\, (X-1)(X+1)}{(2 X - 1)^4} \ee
 and the second order source term is \be
 Src_2 = [-V_2\,-2\, \log X\, \lamt_1 ]\,\psi_0 +[-V_1 -\lamt_1]\, \psi_1
 ~,
\ee
 where the $2 \log X$ term comes from expanding $r^2$.

By (\ref{solve-lam}) \be
 \lamt_2 = \frac{\left< \psi_0|Src_2
\right>}{\left< \psi_0|\psi_0
 \right>}=-1 ~.\ee
Solving (\ref{pert}) for $\psi_2$ we find \bea
 \psi_2 &=& \frac{1}{(2 X-1)^3}
 -\frac{ \log X + 1}{(2 X-1)^2}
 + \frac{Li_2(1-X)}{2 X-1} ~,
 \eea
 where $Li_2(x) \equiv \sum_{k=1}^{\infty}\, \frac{x^k}{k^2}$ is the
 second poly-logarithm function.

\presub {\it Third order}. At this order we can compute
analytically the correction to the eigenvalue \be
 \lamt_3=2 ~,\ee
 where we used \be
 V_3=-\frac{16\, (X-1)^2\, (2 X+1)}{(2X-1)^5} ~.\ee

Here we stop the perturbative expansion as we did not obtain
analytically the fourth order correction to the eigenvalue.

\vspace{.3cm}

Summarizing the high $d$ section, we found that
 assuming the boundary condition (\ref{highd-bc})
the eigenvalue is given by \be
 \lam = \dh - 1 + \frac{2}{\dh} + \dots \equiv d - 4 + \frac{2}{d} + \dots ~~~.\ee

\section{Low $d$}
\label{lowDsection}

In the low $d$ limit the analysis runs in parallel with the previous
high $d$ analysis, but some features differ. We shall see that the
matching problem still reduces to the near horizon region, but here
we can determine analytically only the order of leading behavior and
we need to resort to numerical analysis even in order to determine
the corresponding leading coefficient.

Since we are taking the limit $d-3 \to 0$ it will be convenient to
use another notation for $\dh$ \be
 \eps:= \dh \equiv d-3 ~.\ee

\begin{figure}[t!] \centering \noindent
\includegraphics[width=5cm,angle=-90]{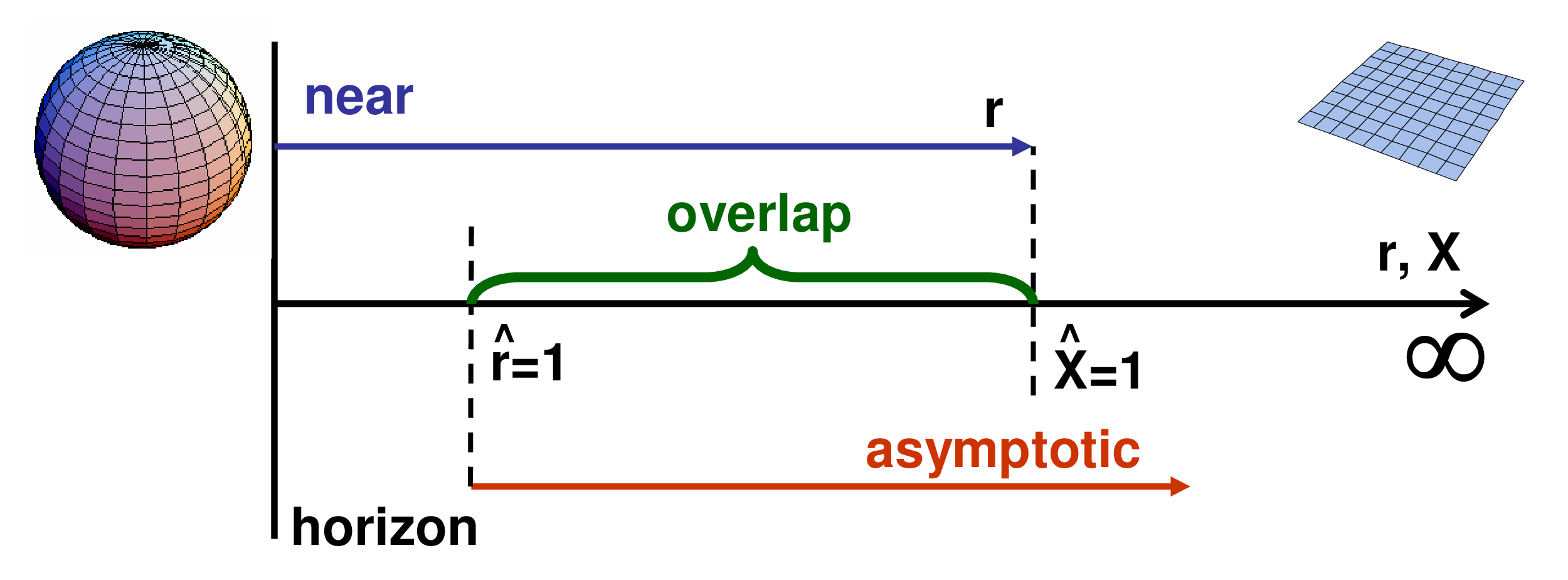}
\caption[]{The regions along the radial direction at low dimension.}
\label{low-figure}
\end{figure}

Unlike the large $d$ limit, here $\xh \ll \rh$, and hence $\rh=1$
occurs much before $\xh=1$ on the $r,X$ axis (see figure
\ref{low-figure}). Accordingly we can define two regions \bi
 \item Near horizon $\xh \ll 1$
 \item Asymptotic $\rh \gg 1$
 \ei
and again the two regions have an overlap which increases in size
as the low $d$ limit is taken, allowing to employ the method of
matched asymptotic expansion.

\presub {\bf Asymptotic region and boundary conditions for  the
near horizon region}.
 In the asymptotic region $\xh$ is a good coordinate,
although $r \gg X$ would continue to be useful. In this region \be
 -V(r) = \frac{2 \eps^3\, (2+\eps)}{r^2\, \(2(1+\eps)\, \xh +
 \eps\)^2} \le \frac{\eps^3}{ r^2\, \xh^2} \ll \lambda \ee
  and hence the potential can be neglected
(in the last inequality it should be borne in mind that
$\lambda={\cal O}(\eps)$ as we shall see later in this section).

On the other hand the function $f(r) = 1-r^{-\eps} = \xh/(1+\xh)$
varies in this region: for large $\xh$ we have $f \simeq 1$ while
for small $\xh$ we have $f \simeq \xh \simeq \eps\, \log(r)$. Yet,
throughout the asymptotic region the leading expression for the
solutions is \be
 \psi = A\, \exp(\sqrt{\lam}\, r) + B\, \exp(-\sqrt{\lam}\, r) ~.\ee
Therefore regularity at infinity requires choosing $A=0$.
Again we shall make a working assumption that this boundary
condition is correct to all orders which we compute and we view
the good agreement with the numerical results as a confirmation.

\presub {\bf The eigenvalue problem}. In the near horizon region
$r$ is a good coordinate. The master equation reads (\ref{master})
\bea
 & & \[-\frac{1}{r^{1+\eps}}\, \del_r\, f(r)\, r^{1+\eps}\, \del_r + V(r)
 \] \psi = -\lam\, \psi \non
 f &=& 1 - \frac{1}{r^{\eps}} \non
 V &=& -\frac{2 \eps^3\, (2+\eps)}{r^2\, \(2(1+\eps)\, (r^{\eps}-1) +
 \eps\)^2} ~~.
 \eea

For fixed $r$ we have $r^{\eps}-1=\exp(\eps\, \log(r)) - 1 =\eps\,
\log(r) + \dots$. Hence for small $\eps$ we may approximate $f$ by
\be f =  \eps\, \log(r) + \dots ~~~. \ee
 In order to have an expression for $V(r)$ in this limit we
expand the following expression from its denominator \be
 Dn_V := 2(1+\eps)\, (r^{\eps}-1) + \eps = \eps(2 \log(r) +1) + \dots ~~~.\ee

Altogether the zeroth order eigenvalue problem is \be
\[ -\frac{1}{r}\, \del_r\, r\, \log(r)\, \del_r -\frac{4}{r^2\,
(2\log(r)+1)^2} \] \psi = -\lamb\, \psi ~, \label{lowd-0th} \ee
 where in this section the rescaled eigenvalue is defined by \be
 \lamb := \lam/\eps ~,\ee
 which is different from the high $d$ definition. The scalar
 product which makes this operator Hermitian is \be
 \left< \phi|\psi \right> = \int_0^\infty \phi^*(r)\,
 \psi(r)\, r dr ~.
\ee

Curiously for $\lamb=0$ there is again an exact solution \be
 \psi = \frac{1-2 \log(r)}{1+2 \log(r)} ~.\ee
It is regular at both $r=1$ and $r=\infty$ but non-normalizable.
For us it is important that it has a single node thereby
indicating the existence of a unique bound state.\footnote{Here
normalizability is not necessary. Imagine increasing $\lam$ while
keeping regularity at the horizon. During this process $\psi(r)$
increases and the node approaches $\infty$. The $\lam$ where the
node becomes infinite is the sought for ground state.} Unlike high
$d$ here the marginally bound state is found not to become bound
for $\eps>0$ -- which is fortunate, since we are expecting only a
single bound state.

So far we have shown the existence of a single bound state whose
eigenvalue scales as $\lam = c_1\, \eps + \dots$ where $c_1$ is some
positive constant. In particular \be
 \lam(3d)=0 ~.\ee
We were not able to solve for $c_1$ analytically. But a numerical
analysis of (\ref{lowd-0th}) yields
 $c_1 = .71515$.

One can proceed to higher orders in perturbation theory
$\lam=c_1\, \eps + c_2\, \eps^2 + \dots$  Clearly this needs to be
done numerically as even the zeroth order eigenvalue and
eigenfunction are known only numerically. Doing so we obtained
$c_2 = 0.0627$.




\vspace{.3cm}

In summary the low $d$ expansion is \bea
 \lam &=& c_1\, (d-3) + c_2\, (d-3)^2 + \dots \non 
 c_1 &=& 0.71515 \non
 c_2  &=& 0.0627 ~.\eea

\section{Numerical data and interpolations}
\label{InterSection}

We need high-precision numerical data for $\lam(d)$ in order to
confirm our perturbative results and in order to evaluate the
success of various interpolating functions based on these
expansions. To that end we computed $\lam(d)$ to 5-6 digits of
precision by fairly standard numerical procedures and the results
are collected in a table is appendix \ref{numeric}.

We first confirm that the expansions which we derived
(\ref{highd-sum},\ref{lowd-sum}) are consistent with the data. The
method is straightforward: to test an order $k$ Taylor expansion
around $r_0$ assuming order $k-1$ was confirmed already, we deduct
the partial sum up to order $k-1$ from the data, divide by the
small parameter to power $k$ and seek to identify a limit as $r
\to r_0$ (extrapolations may be used to increase the accuracy of
the limit). Using this method we confirm
 the first two coefficients in
(\ref{highd-sum},\ref{lowd-sum}),
 thereby giving us added confidence in the validity of our
 boundary conditions.
The agreement with the $2/d$ term in the high $d$ expansion is
less clear. The remainder function which estimates the coefficient
of the $1/d$ term is indeed very close to 2 for $12 \lesssim d
\lesssim 21$ but then for higher $d$ it increases and gets as high
as 2.5 and 3. We interpret this increase to be due to either the
diminishing numerical accuracy for high $d$ caused by the
multiplication by $d$ or to a correction to the boundary condition
(\ref{highd-bc}) at this order.


Next we would like to exploit our control over the function at the
two boundaries in order to construct an interpolating function which
would be a reasonable approximation across the whole range $3 \le d
< \infty$. We shall require the function to coincide with the
derived Taylor expansion on both side up to a prescribed order. A
popular and simple kind of interpolating function is provided by the
class of rational functions. Note that while the Pad\'{e}
approximation also employs rational functions, it is usually fitted
to a Taylor expansion at a single point and is therefore an
\emph{extrapolation} which is much less reliable than the current
\emph{interpolation}.

It is observed that $\lam/\dh$ is a slowly varying function. A
simple possibility for a rational interpolating function is \be
 \frac{\lam}{\dh} \simeq \frac{a_0+a_1/\dh}{b_0+b_1/\dh} ~. \ee
 This equation has 3 unknowns (an overall multiplication of all constants does
not change the function) and thus can be made to fit 2 of the
coefficients at high $d$ and the single $c_1$ coefficient of low
 $d$. The resulting rational function is \be
 \Lam(d) = \dh ~ \frac{(1-c_1)\, \dh + c_1}{(1-c_1)\, \dh + 1}
 \label{interpol-a}
 ~. \ee

The approximation $\Lam(d)$ agrees quite well with the data for
$\lam(d)$, as can be seen from figures
\ref{compare1},\ref{compare2}.

\EPSFIGURE{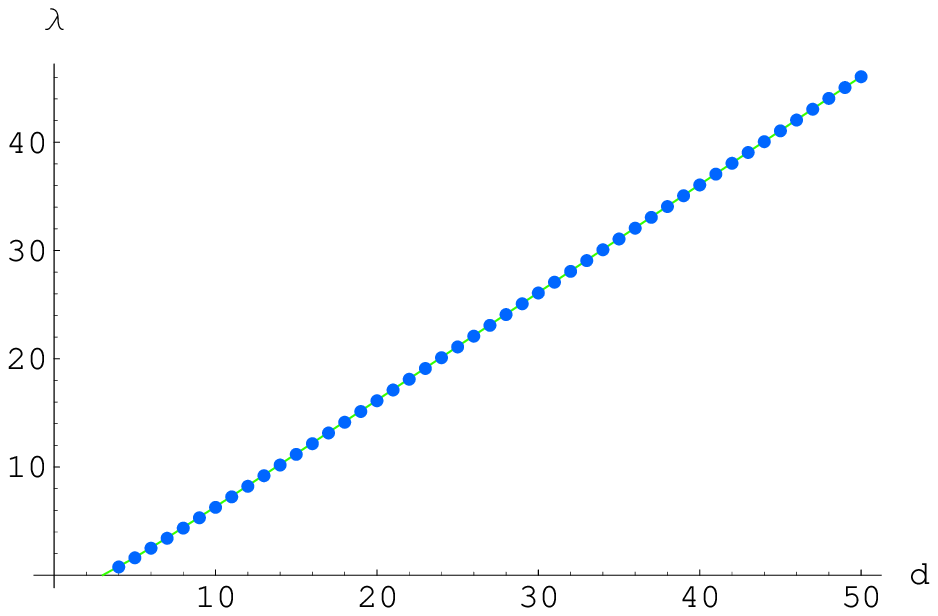}{The negative eigenvalue $\lam$ as a
function of $d$ the space-time dimension. The blue dots represent
the numerical data (see appendix \ref{numeric}) while the solid
green line represents the interpolation (\ref{interpol-a}) with
$c_1=0.71515$. \label{compare1}}

 \EPSFIGURE{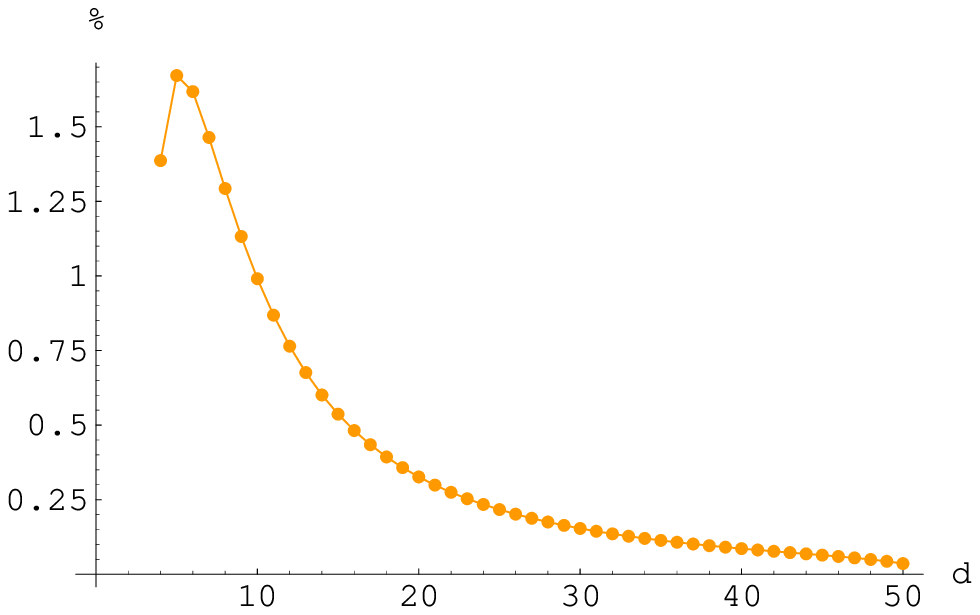}{The
error of interpolation in percent as a function of $d$
(the data is represented by the heavy dots and the solid curve is
intended merely to guide the eye).
It is seen that the maximal error is less than $1.7$\% and for $d
\ge 10$ it is even less that 1\%. \label{compare2}}

We also attempted interpolation with rational functions of higher
order, and we found that not only does their approximation error
not improve relative to (\ref{interpol-a})  but it is actually even
larger.

\subsubsection*{Acknowledgements}

We thank Michael Smolkin for discussions.

This research is supported by The Israel Science Foundation grant
no 607/05, DIP grant H.52, EU grant MRTN-CT-2004-512194 and the
Einstein Center at the Hebrew University. U.M. is supported by a
Grant for the 21st Century COE Program (Holistic Research and
Education Center for Physics Self-Organization Systems) at Waseda
University.

\appendix


\section{High precision numerical values for $\lam(d)$}
\label{numeric}

In this section we present in table \ref{table-lam} high precision
numerical results for $\lam(d)$ which is defined by the eigenvalue
problem (\ref{master}). For more details on the method of
calculation see the Mathematica notebook \cite{MathUmpei}.

\begin{table}[h!]
\centering \noindent
\begin{tabular}{|c||c|c|c|c|c|}\hline
 d &  4 & 5 & \multicolumn{3}{c|}{} \\
 $\lam$  &0.76766 &1.61015 & \multicolumn{3}{c|}{} \\ \hline \hline
  d  &   6&7&8&9&10  \\
$\lam$   &2.49879& 3.41739& 4.35618& 5.30901& 6.27189
\\ \hline \hline
  d& 11& 12 & 13& 14& 15 \\
$\lam$  & 7.24210& 8.21779& 9.19766& 10.1808 & 11.1664
\\ \hline \hline
 d& 16& 17& 18& 19& 20 \\
 $\lam$ & 12.1541& 13.1434& 14.1341& 15.1259& 16.1186 \\ \hline \hline
 d& 22& 24& 26& 28& 30 \\
 $\lam$ & 
  18.1062 &
  20.0962 &
  22.0879 &
  24.0810 &
  26.0750 \\ \hline \hline
 d& 32& 34& 36& 38& 40 \\
 $\lam$ & 
 28.0700 &
 30.0656&
 32.0618&
 34.0585&
 36.0557 \\ \hline \hline
 d& 42& 44& 46& 48& 50 \\
 $\lam$ & 
 38.0533&
 40.0515&
 42.0504&
 44.0505&
 46.0530 \\ \hline
\end{tabular}
\caption[]{Numerically computed high-precision negative eigenvalues
  $\lam$ of (\ref{master}) (in units of $r_0^{-2}$). }
\label{table-lam}
\end{table}
%


\end{document}